\documentstyle[12pt]{article}

\begin{document}


\title{Classical Underpinnings of Gravitationally Induced Quantum 
Interference\footnote
{UCONN 96-08, October 1996}}

\author{\normalsize{Philip D. Mannheim} \\
\normalsize{Department of Physics,
University of Connecticut, Storrs, CT 06269} \\ 
\normalsize{mannheim@uconnvm.uconn.edu} \\}

\date{}

\maketitle

\begin{abstract}
We show that the gravitational modification of the phase of a 
neutron beam (the COW experiment) has a classical origin, being
due to the time delay which classical particles experience in 
traversing a background gravitational field. Similarly, we show 
that classical light waves also undergo a phase shift in 
traversing a gravitational field. We show that the COW experiment 
respects the equivalence principle even in the presence of quantum 
mechanics. 
\end{abstract}

In a landmark series of experiments \cite{Overhauser1974}, \cite{Colella1975} 
Colella, Overhauser and Werner (COW) and their subsequent collaborators (see 
e.g. Refs. \cite{Greenberger1979}, \cite{Werner1994} for overviews) detected the 
modification of the phase of a neutron beam as it traverses the earth's 
gravitational field, to thus realize the first experiment which involved both 
quantum mechanics and gravity. A typical generic experimental set up is shown in 
the schematic Fig. (1) in which a neutron beam from a reactor is split by 
Bragg or Laue scattering at point $A$ into a horizontal beam $AB$ and a vertical 
beam $AC$ (we take the Bragg angle to be $45^{\circ}$ for simplicity and 
illustrative convenience in the following), with the subsequent scatterings at 
$B$ and $C$ then producing beams which Bragg scatter again at $D$, after which 
they are then detected. If the neutrons arrive at $A$ with velocity 
$v_0$ (typically of order $2\times 10^5$ cm sec$^{-1}$) and $ABCD$ is a square 
of side $H$, then the phase difference $\phi_{COW}=\phi_{ACD}-\phi_{ABD}$ 
is given by $-mgH^2/\hbar v_0$ to lowest order in the acceleration $g$ due to 
gravity \cite{Overhauser1974}, and is actually observable despite the weakness 
of gravity, since even though $\int \bar{p} \cdot d\bar{r}$ only differs by the 
very small amount $m(v_{CD}-v_{AB})H=-mgH^2/v_0$ between the $CD$ and $AB$ 
paths, nonetheless this quantity is not small compared to Planck's constant, to 
thus give an observable fringe shift even for $H$ as small as a few centimeters.

The detected COW phase is extremely intriguing for two reasons. First, it shows 
that it is possible to distinguish between different paths which have common end 
points, with the explicit global ordering in which the horizontal and vertical 
sections are traversed leading to observable consequences. And second, it yields 
an answer which explicitly depends on the mass of the neutron even while the 
classical neutron trajectories (viz. the ones explicitly followed by the centers 
of the wave packets of the quantum mechanical neutron beam) of course do not. 
The COW result thus invites consideration of whether the detected ordering is 
possibly a topological effect typical of quantum mechanics, and of whether 
quantum mechanics actually respects the equivalence principle.\footnote{Possible 
implications of the mass dependence of the COW phase on another related issue,
viz. quantum mechanics and the action and reaction principle, are explored in 
Ref. \cite{Anandan1995}.} As we shall see, the ordering effect is in fact
already present in the motion of classical particles in gravitational fields,
and even in the propagation of classical waves in the same background, with
this latter feature enabling us to establish below that the mass dependence of
the neutron beam COW phase is purely kinematic with the equivalence principle
then not being affected. 

To address these issues specifically we have found it convenient to carefully
follow the neutron as it traverses the interferometer, to find that the two
beams do not in fact arrive at the same point $D$ or even at the same time, with
this spatial offset and time delay not only producing the interference effect,
but also being present in the underlying classical theory. Quantum mechanics
thus does not cause the time delay, rather it only serves to make it observable.
Since gravity is a relativistic theory we shall need to introduce curvature
(which we do below), but we have found it more instructive to consider the 
non-relativistic limit first. Since we can treat the neutron beams as rays, 
their motions round the $ABCD$ loop can be treated purely classically between 
the various scatterings. Moreover, the various scatterings themselves 
at $A$, $B$, $C$ and $D$ introduce no additional phases, are energy conserving, 
and give angles of reflection equal to the angles of incidence.\footnote{This is 
not completely obvious since if the neutron slows down a little or is deflected 
a little, then at each subsequent scattering it is not quite incident with a 
wavelength and angle which satisfy a Bragg peak condition. However, consider 
a crystal for which Bragg scattering would occur for the pair of vectors 
$\bar{k}_i$ and $\bar{k}_f$, viz. equal length vectors which obey 
$\bar{k}_i -\bar{k}_f=\bar{G}$ where $\bar{G}$ is a reciprocal lattice vector  
which imparts a momentum transfer perpendicular to the scattering surface. Now
instead allow $\bar{k}_i$ to impinge upon a crystal with additional momentum
$\Delta \bar{k}_i$ with the outgoing $\bar{k}_f$ beam then emerging with  
additional momentum $\Delta \bar{k}_f$. Since each such scattering is 
coplanar the most general such momentum modification can conveniently be 
written as $\Delta \bar{k}_i =\alpha_i \bar{k}_i +\alpha_f \bar{k}_f$, 
$\Delta \bar{k}_f =\beta_i \bar{k}_i +\beta_f \bar{k}_f$, with energy 
conservation yielding $\beta_i=\alpha_f$, $\beta_f=\alpha_i$ since there is no 
energy transfer to the crystal. Consequently, the momentum transfer at a 
scattering is now given by $\bar{k}_i +\Delta \bar{k}_i -\bar{k}_f-\Delta 
\bar{k}_f =(1+\alpha_i - \alpha_f)\bar{G}$, and is thus still perpendicular
to the scattering plane, but now no longer right on a diffraction maximum.
The angle of reflection is thus still equal to the angle of incidence, but
with a scattering intensity which is slightly reduced with respect to the 
$\Delta \bar{k}=0$ case. (The only way that we would actually be able to shift 
into a new Bragg peak condition would be to have $\alpha_i-\alpha_f =0$, a 
condition which only occurs when each $\Delta \bar{k}$ is parallel to the 
scattering surface.)} Thus the entire motion of the neutron is the same as that 
of a classical macroscopic particle (the neutron spin plays no explicit role in 
the COW experiment, so we treat the neutrons as spinless) which undergoes 
classical mirror reflections.

The neutron which goes up vertically from $A$ to $C$ arrives at $C$ with a 
velocity $(0,v_0-gH/v_0)$. The neutron $AC$ travel time is $t(AC)$ and the 
classical action $S=\int (\bar{p} \cdot d\bar{r}-E_0dt)$ ($E_0=mv_0^2/2$) 
undergoes a change $S(AC)$ where
\begin{equation}
t(AC)=H/v_0+gH^2/2v_0^3~~~,~~~S(AC)=mv_0H-mgH^2/2v_0 -E_0t(AC)
\label{(1)}
\end{equation}
\noindent On scattering at $C$ the neutron is then reflected so that it starts 
off toward $D$ with a velocity $(v_0-gH/v_0,0)$. On its flight it dips slightly 
to arrive at the next scattering surface at the point $D_1$ with 
coordinates $(H-gH^2/2v_0^2,H-gH^2/2v_0^2)$, so that there is a change in the 
end point of the motion which is first order in $g$ and thus relevant to our
discussion. At $D_1$ the neutron has a velocity $(v_0-gH/v_0,-gH/v_0)$, 
with the $CD_1$ segment taking a time $t(CD_1)$ and contributing an amount 
$S(CD_1)$ to the classical action where
\begin{equation}
t(CD_1)=H/v_0+gH^2/2v_0^3~~~,~~~S(CD_1)=mv_0H-3mgH^2/2v_0 -E_0t(CD_1)
\label{(2)}
\end{equation}
\noindent
The neutron which starts horizontally from $A$ arrives not at $B$ but at the 
point $B_1$ with coordinates $(H-gH^2/2v_0^2,-gH^2/2v_0^2)$ 
and with a velocity $(v_0,-gH/v_0)$. The $AB_1$ segment takes a time
$t(AB_1)$ and the action changes by $S(AB_1)$ where
\begin{equation}
t(AB_1)=H/v_0-gH^2/2v_0^3~~~,~~~S(AB_1)=mv_0H-mgH^2/2v_0 -E_0t(AB_1)
\label{(3)}
\end{equation}
\noindent
After scattering at $B_1$ the neutron sets off toward $D$ with velocity
$(-gH/v_0,v_0)$ and arrives not at $D$ or $D_1$ but rather at the point
$D_2$ with coordinates $(H-3gH^2/2v_0^2,H-3gH^2/2v_0^2)$, and
reaches there with velocity $(-gH/v_0,v_0-gH/v_0)$. The 
$B_1D_2$ segment takes a time
$t(B_1D_2)$ and the action changes by $S(B_1D_2)$ where
\begin{equation}
t(B_1D_2)=H/v_0-gH^2/2v_0^3~~~,~~~S(B_1D_2)=mv_0H-3mgH^2/2v_0 -E_0t(B_1D_2)
\label{(4)}
\end{equation}
\noindent
As regards the neutron's path around the loop, we see that the small vertical 
$gH^2/2v_0^2$ dip during each of the two horizontal legs causes each neutron to 
travel a distance $gH^2/2v_0^2$ less in the horizontal than it would have done 
in the absence of gravity, to thus provide a first order in $g$ modification to 
$\int \bar{p} \cdot d\bar{r}$ in each of these legs, even while these same 
vertical dips themselves only contribute to the action in second order. However, 
for the two horizontal sections, each leg is shortened by the same amount in 
the horizontal, so that the difference in $\int \bar{p} \cdot d\bar{r}$ between 
the $CD_1$ and $AB_1$ legs still takes the value $m(v_{CD}-v_{AB})H=-mgH^2/v_0$ 
quoted earlier. As regards the two vertical legs, we note that even though the 
$AC$ leg is completely in the vertical, since the neutron starts the $B_1D_2$ 
leg with a small horizontal velocity, during this leg the neutron changes its 
horizontal coordinate by an amount $gH^2/v_0^2$, thereby causing it to reach 
$D_2$ after having also traveled a distance $gH^2/v_0^2$ less in the vertical 
than it would travel in the $AC$ leg. Consequently, there is both a spatial 
offset $(gH^2/v_0^2,~gH^2/v_0^2)$ between $D_1$ and $D_2$, and a time  
delay $t(AC)+t(CD_1)-t(AB_1)-t(B_1D_2)= 2gH^2/v_0^3$ between the arrival of the 
two beams, with the integral $\int \bar{p} \cdot d\bar{r}$ thus not taking 
the same value in each of the two vertical legs. However, our calculation shows 
that all the these modifications actually compensate in the overall loop with 
there being no difference in $\int \bar{p} \cdot d\bar{r}$ between the $ACD_1$ 
and $AB_1D_2$ paths. However, since there is a time delay, there is still a net 
change in the action, given by $S(AC)+S(CD_1)-S(AB_1)-S(B_1D_2)=-mgH^2/v_0$. 
However, we cannot identify this quantity with the COW phase 
$\hbar \Delta \phi_{COW}$  since the beams have not interfered due to the 
spatial offset between $D_1$ and $D_2$.

Before discussing the issue of this spatial offset, it is instructive to ask
where the neutrons would have met had there been no third crystal at $D$ to get 
in the way. Explicit calculation shows that they would in fact have met at the 
asymmetric point $D_3$ with coordinates 
$(H-3gH^2/2v_0^2,H-gH^2/2v_0^2)$
with the $CD_3$ and $B_1D_3$
segments taking times $t(CD_3)$ and $t(B_1D_3)$ while yielding action 
changes $S(CD_3)$ and $S(B_1D_3)$, where
\begin{eqnarray}
t(CD_3)=H/v_0-gH^2/2v_0^3~~~,~~~S(CD_3)=mv_0H-5mgH^2/2v_0 -
E_0t(CD_3)
\nonumber \\
t(B_1D_3)=H/v_0+gH^2/2v_0^3~~~,~~~S(B_1D_3)=mv_0H-mgH^2/2v_0 -
E_0t(B_1D_3)
\label{(5)}
\end{eqnarray}
\noindent 
The neutrons would thus meet at $D_3$ without any time delay, and with 
$S(AC)+S(CD_3)-S(AB_1)-S(B_1D_3)=-2mgH^2/v_0$. We thus see that for purely 
classical particles reflecting off mirrors at $B_1$ and $C$ the quantity 
$\int \bar{p} \cdot d\bar{r}$ evaluates differently for the two paths $ACD_3$ 
and $AB_1D_3$. This is a thus a global, path dependent effect in purely 
classical mechanics in a background classical gravitational field which is 
completely independent of quantum mechanics.\footnote{Since the momentum 
$\bar{p}$ is the vector derivative $\bar{\nabla} S$ of the stationary classical 
action, for any trajectory which obeys the equations of motion the integral 
$\int \bar{p} \cdot d\bar{r}$ then only depends on its endpoints (in a manner 
which in general depends on the explicit $\bar{\nabla} S$ direction in which 
each endpoint is approached.) Consequently, it is at first a little surprising 
that $\int \bar{p} \cdot d\bar{r}$ could evaluate differently for the $ACD_3$ 
and $AB_1D_3$ paths, since these paths do share common endpoints. However, 
neither one of these paths is the one which is the stationary one for the 
motion which starts at $A$ and finishes at $D_3$ (viz. a parabola close to the 
straight line $AD_3$); and thus even while the $ACD_3$ path for instance is 
composed of the two subpaths $AC$ and $CD_3$ each one of which is stationary 
between its own endpoints, nonetheless 
the sum of the two subpaths is not stationary between the overall $A$ and $D_3$ 
endpoints, with $\bar{p}$ then not being expressible as a total derivative for 
the whole $ACD_3$ (or $AB_1D_3$) motion. In passing, we note that the fact that 
the sum of two stationary paths is not necessarily stationary between the 
overall endpoints is also the origin (see e.g. \cite{Mannheim1983}) of the 
existence of the many paths of the Feynman path integral description of quantum 
mechanics.} However, since the classical action is not observable in classical 
mechanics, it is only in the presence of quantum mechanics that phase 
differences become observable. (In classical mechanics what is observable is 
that the neutrons meet at $D_3$ rather than on the $AD$ axis.)

Returning now to the COW experiment itself, in order to understand the 
implications of the time and spatial offsets between $D_1$ and 
$D_2$, it is instructive to consider the Young double slit 
experiment with purely classical light. As shown in Fig. (2), light from a
source $S$ goes through slits $Q$ and $R$ to form an interference pattern
at points such as $P$, with the distance $\Delta x=QT$ representing the 
difference in path length between the two beams. Given this path difference, 
the phase difference between the two beams is usually identified as $k\Delta x$, 
from which an interference pattern is then readily calculated. However, because 
of this path difference, the $SQP$ ray takes the extra time $\Delta t 
=\Delta x/c$ to get to $P$, to thus give a net change in the phase of the $SQP$ 
beam of $k\Delta x-\omega \Delta t$ which actually vanishes for light rays. 
The relative phase of the two light rays in the double slit experiment thus does 
not change at all as the two beams traverse the interferometer. However, because 
of the time delay, the $SRP$ beam actually interferes with an $SQP$ beam which 
had left the source a time $\Delta t$ earlier. Thus if the source is coherent 
over these time scales, the $SQP$ beam carries an additional  
$+\omega \Delta t$ phase from the very outset. This phase then cancels the 
$-\omega \Delta t$ phase it acquires during the propagation to $P$ (a 
cancellation which clearly also occurs for quantum mechanical matter waves 
moving with velocities less than the velocity of light), leaving just 
$k\Delta x$ as the final observable phase difference, a quantity
which is non-zero only if there is in fact a time delay. We thus see that the 
double slit device itself actually produces no phase change for light. 
Rather, the choice of point $P$ on the screen is a choice which selects which
time delays at the source are relevant at each $P$, with the general 
interference pattern thus not only always involving the time delay at the 
source, but also in fact always requiring one. 

With this in mind, we now see that we also need to monitor the time delay of 
the neutron in the COW experiment, However, since the total energy of the 
neutron does not change as it goes through the interferometer, the time delay 
contribution will still drop out of the final phase shift expression 
(explicitly but not implicitly). However, for the COW experiment we noted above 
that as well as a time delay between the $ACD_1$ and $AB_1D_2$ paths, there was 
also a spatial offset. Consequently, the $AB_1D_2$ path interferes not with the 
$ACD_1$ path, but rather with the indicated offset $A_1C_1D_2$ path, a very 
close by path which in fact is found to lie a distance $gH^2/v_0^2$ vertically 
below $AB$, an offset distance which is within the resolution of the beam. The 
evaluation of the phase shift is then exactly as before with $S(A_1C_1D_2)$ 
taking the exact same value as $S(ACD_1)$ to lowest order in $g$. Now we noted 
above that all the $\int \bar{p} \cdot d\bar{r}$ contributions actually cancel 
for this particular set of paths. However, because of the spatial offset 
between $D_1$ and $D_2$, the $AB_1D_2$ path beam has to travel an extra 
horizontal distance $A_2A=gH^2/v_0^2$ to first get to the interferometer 
(to therefore provide an analog to the distance $\Delta x=QT$ in the double 
slit experiment, with $A_1$ and $A_2$ acting just like the pair of slits $Q$ 
and $R$). Now in traveling this extra $A_2A$ distance this beam actually 
acquires yet another time delay $t(A_2A)$, to therefore impose yet another 
relative phase condition at the source which then identically cancels the 
associated $-E_0t(A_2A)$ change in the action. Moreover, in traveling this 
extra $A_2A$ the integral $\int \bar{p} \cdot d\bar{r}$ acquires yet one more 
contribution $mgH^2/v_0$, and this term then emerges as the only contribution 
in the entire circuit which is not canceled. Consequently, we obtain 
$\Delta \phi_{COW}=-mgH^2/\hbar v_0$ as the final observable net 
COW phase shift.\footnote{While we obtain the same answer
for $\Delta \phi_{COW}$ as previous authors, nonetheless our derivation
seems to be somewhat different from the previously published ones.}

Turning now to a fully covariant analysis, we need to look at solutions to
the Klein-Gordon equation $\phi^\mu _{\phantom{\mu};\mu}-(mc/\hbar)^2\phi=0$ 
($\phi^{\mu}$ denotes $\partial\phi/\partial x_{\mu}$) in generic background 
fields of the form $d\tau^2=B(r)c^2dt^2-dr^2/B(r)-r^2d\Omega$ where 
$B(r)=1-2MG/c^2r$. Firstly we note that the non-relativistic reduction of this 
Klein-Gordon equation is straightforward, with the substitution 
$\phi=$exp$(-imc^2t/\hbar)\psi$ then yielding 
\begin{equation}
i\hbar{\partial \psi \over \partial t}+{\hbar^2 \over 2m} \nabla ^2 \psi=
{mc^2 \over 2}\left( B(r)-1 \right)=-{mMG \over r}
\label{(6)}
\end{equation}
for slowly moving particles. We thus see that the inertial mass $m$ which is
defined via the starting Klein-Gordon equation thus also serves as the passive
gravitational mass which serves to couple massive particles to gravity, so
that the particle modes associated with the quantization of the Klein-Gordon
field thus automatically obey the equivalence principle, precisely because of 
quantum mechanics in fact.\footnote{We note that since the starting covariant 
Klein-Gordon equation only possesses one intrinsic mass scale, it would appear 
that the equivalence principle has to emerge. However, there is actually a 
hidden assumption in the use of the covariant Klein-Gordon equation (see 
\cite{Mannheim1993}) since it is not the most general equation in curved space
which can reduce to the flat Klein-Gordon equation in flat space. Rather, the
curved space equation could also possess additional explicitly curvature 
dependent terms, terms which would then vanish in the flat space limit, but
which would modify the particle's coupling to gravity in curved space in a
potentially equivalence principle violating manner. Now such possible terms 
(which would of course have to produce effects of order less than $10^{-11}$ so 
as not to spoil the Eotvos result) are simply ignored in standard gravity 
without any apparent justification as far as we can tell, with the standard 
gravitational phenomenology only in fact following in their assumed absence. 
Thus in passing it is of interest to note that any such possible additional 
curvature dependent terms are not in fact allowed to appear \cite{Mannheim1993} 
in the conformal invariant gravitational alternative currently being considered 
by Mannheim and Kazanas (see e.g. \cite{Mannheim1994}).} 

As regards the covariant Klein-Gordon equation, it is convenient to make the 
substitution $\phi(x)=$exp$(iS(x)/\hbar)$, so that the phase then obeys
$S^{\mu}S_{\mu}+m^2c^2=i\hbar S^{\mu}_{\phantom{\mu};\mu}$. In the eikonal or 
ray approximation the $i\hbar S^{\mu}_{\phantom{\mu};\mu}$ term can be dropped, 
so that the phase $S(x)$ is then seen to obey the classical Hamilton-Jacobi 
equation $S^{\mu}S_{\mu}+m^2c^2=0$, an equation whose solution is the stationary 
classical action between relevant end points. We thus establish that in the 
eikonal approximation the phase of the wave function of a material particle 
is in fact the classical action just as required for the discussion of the COW 
experiment we gave earlier. In the eikonal approximation we can also identify 
$S^{\mu}$ as the momentum $p^{\mu}=mcdx^{\mu}/d\tau$, so that we can set 
$S(x)=\int p_{\mu}dx^{\mu}$; with the covariant differentiation of the 
Hamilton-Jacobi equation then yielding \cite{Mannheim1993} 
$p^{\mu}p^{\nu}_{\phantom{\nu};\mu}=0$, which we recognize as the massive 
particle geodesic equation.

In order to actually calculate the geodesics in the gravitational field of
the earth it is convenient to rewrite the Schwarzschild metric in terms
of a Cartesian coordinate system $x=r sin \theta cos \phi,~y=r sin \theta sin 
\phi,~z=r cos \theta -R$ erected at a point on the surface of
the earth. With $z$ being normal to the earth's surface, to lowest order in 
$x/R,~y/R,~z/R,~MG/c^2R$ (where $M$ is the mass of the earth and $R$ its 
radius) the Schwarzschild metric is then found \cite{Moreau1994} to take the 
form
\begin{equation}
d\tau^2=f(z)c^2dt^2-dx^2-dy^2-dz^2/f(z)
\label{(7)}
\end{equation}
where $f(z)=1-2MG/c^2R+2gz/c^2$ and where $g$ denotes $MG/R^2$. For this 
metric the non-relativistic geodesics for material particles are given by 
$\ddot{x}=0,~\ddot{y}=0,~\ddot{z}=-g$, to thus enable us to completely justify 
our earlier non-relativistic calculation.\footnote{Since the metric of Eq. 
(\ref{(7)}) yields a uniform gravitational acceleration, it must therefore be 
flat, and indeed it can readily be transformed into a flat Cartesian metric 
$d\tau^2=c^2dt^{\prime 2} -dx^2-dy^2-dz^{\prime 2}$ via the coordinate 
transformation \cite{Greenberger1979} $ct^{\prime} 
=c^2sinh(gt/c) f^{1/2}(z)/g$, $z^{\prime}=c^2(cosh(gt/c)f^{1/2}(z)
-f^{1/2}(0))/g$ under which the small $(z,t)$ 
region transforms into the small $(z^{\prime},t^{\prime})$ region. However, 
under such a transformation the time delay in the COW experiment would not be 
removed (the time delay between the two beams at the source is a timelike vector 
which remains timelike under any coordinate transformation), with the fact of a 
time delay thus being a covariant indicator for the COW effect. Similarly,
since the action is the covariant scalar $\int p_{\mu}dx^{\mu}$, its value
in any trajectory must remain unchanged under this transformation to flat 
Cartesian coordinates. Now while there is no explicit reference to $g$ in flat 
coordinates, nonetheless, in an accelerated coordinate frame in flat space
there is still an implicit dependence on $g$, since points such as $D$ will 
accelerate upward with acceleration $g$ causing the beam from $C$ to arrive at 
the $g$ dependent $D_1$, to thus expressly show the need for first order 
changes in the end points of the interferometer legs.}

As regards the purely classical, massless case, on defining  
$\phi(x)=$exp$(iT(x))$, we can this time identify the eikonal phase derivative 
$T^{\mu}$ with the wave number $k^{\mu}=dx^{\mu}/dq$ where $q$ is a convenient
affine parameter which can be used to measure distances along trajectories. In 
the massless case the Hamilton-Jacobi equation takes the form
\begin{equation}
f(z)(k^0)^2-(k^1)^2-(k^2)^2-(k^3)^2/f(z)=0
\label{(8)}
\end{equation}
and again yields the requisite massless particle geodesic equation 
$k^{\mu}k^{\nu}_{\phantom{\nu};\mu}=0$ just as in the massive case. In the 
Schwarzschild geometry of interest these massless geodesic equations are found 
to take the form 
\begin{eqnarray}
k^0=cdt/dq=\alpha_0/f(z)~~~,~~~k^1=dx/dq=\alpha_1
\nonumber \\
k^2=dy/dq=\alpha_2~~~,
~~~k^3=dz/dq=(\alpha_0^2-(\alpha_1^2+\alpha_2^2)f(z))^{1/2}
\label{(9)}
\end{eqnarray}
where the $\alpha_i$ are integration constants. From these equations we 
recognize the existence of the gravitational frequency shift (since $k^0$ 
depends on $z$), the gravitational time delay ($dz/dt$ depends on $z$), and the 
gravitational bending of light ($dz/dx$ depends on $z$ if $\alpha_1\neq 0$).
All of these effects are found to be of relevance for the motion of a 
classical light wave around the $ABCD$ interferometer loop,\footnote{As regards 
the Bragg scattering rules for light in curved space, it is shown in 
\cite{Mannheim1996a} that a wave undergoes no change in the magnitude of 
$k^0$ in a Bragg scattering, that it emerges with an angle of reflection 
equal to the angle of incidence, and that the magnitudes of the spatial 
components of the outgoing momentum take whatever values are needed to keep the 
outgoing wave on the light cone of Eq. (\ref{(8)}). Thus, an 
upward going $k^3=\alpha_0$ light wave, for instance, will reflect at $C$ into 
an initially horizontal light wave with momentum $k^1=\alpha_0/f^{1/2}(H)$.} 
with explicit calculation (the details
will be published elsewhere \cite{Mannheim1996a}) then showing that the ensuing 
light rays again follow Fig. (1) around the interferometer. However, even while 
there is still a spatial offset $A_1A_2=gH^2/c^2$ just as before, for light 
neither a time delay nor any net phase shift is found between the $A_1C_1D_2$ 
and $AB_1D_2$ paths. However, again as with the neutron case, the spatial 
offset itself leads to a time delay $A_2A/c$, so that there is still observable 
interference. Then, with $\alpha_0$ replacing $mc/\hbar$ in the normalization 
of the phase shift, we thus find that in traversing the interferometer the two 
beams acquire a final net relative phase shift $-\alpha_0gH^2/c^2$ due entirely
to the $A_2A$ segment alone. On recognizing that $\alpha_0$ is 
the value of $k^0$ at $z=0$ we may set it equal to $2\pi/\lambda$ where 
$\lambda$ is the wavelength of the incident beam, to finally obtain for the 
phase shift $\Delta \phi_{CL}=-2\pi gH^2/\lambda c^2$ where $CL$ denotes 
classical light. Now while $H$ would have to be of the order of $10^5$ cm for 
$\Delta \phi_{CL}$ to actually be detectable in a Bragg scattering 
interferometer of the same sensitivity as the COW experiment,\footnote{While 
this is still sizable for an interferometer, its interest lies in the fact 
that it allows us to detect, in principle at least, the gravitational bending 
of light using laboratory sized distance 
scales rather than solar system sized ones. Thus it would
also be of interest to see what dimension interferometer might serve as a 
gravitational wave detector or be sensitive to any possible neutrino masses in 
neutrino beam interferometry.} nonetheless we can still identify this phase 
shift as an in principle, completely classical effect which reveals the 
intrinsically global nature of classical gravity.\footnote{Given our 
long experience with Newtonian potentials, we are used to the notion that
gravity is local, with the most distant matter making the smallest contribution
at a local point (and none at all if distributed spherically), with the 
non-vanishing of $\Delta \phi_{CL}$ thus exposing a fundamental difference 
between Newtonian and relativistic gravity. Thus it is of some interest to note 
that the potential of conformal gravity \cite{Mannheim1994} takes the form 
$V(r)=-MG/r +\gamma c^2r/2$ where $\gamma$ is a new parameter for sources, to 
show that there are theories of gravity in which even the potentials have 
global aspects, with the most important gravitational contributions at a given 
point then coming from the most distant objects, a very Machian situation. In 
fact, it has very recently been shown \cite{Mannheim1996b} that because of this 
global aspect of conformal gravity, the global 
Hubble flow then explicitly modifies the motion of matter within individual 
galaxies in a way which is then able to account for the observed systematics of 
galactic rotation curves without any apparent need for galactic dark matter.}  

Now that we have obtained $\Delta \phi_{CL}=-2\pi gH^2/\lambda c^2$ it is 
instructive to compare it with $\Delta \phi_{COW}$. If we introduce the neutron 
de Broglie wavelength $\lambda_n = h/mv_0$ we may rewrite 
$\Delta \phi_{COW}$ in the form $-2\pi gH^2/\lambda_n v_0^2$. Comparison with 
$\Delta \phi_{CL}$ thus reveals a beautiful example of wave particle duality, 
with the quantum mechanical matter wave inheriting its interference aspects from 
the behavior of the underlying classical wave. Thus even while 
$\Delta \phi_{COW}$ does depend on the mass of the neutron,\footnote{The reason 
why $\Delta \phi_{COW}$ actually depends on $m$ at all is that even though the 
position of the minimum of the classical action is independent of $m$ (the 
equivalence principle) nonetheless the actual value of the classical action in 
this minimum does depend on $m$ (though only as a purely kinematic overall 
multiplying factor for $-mc\int d\tau$ which does not affect the position of the 
minimum); and even though the value of the classical action is not observable 
classically, nonetheless it is observable quantum mechanically as the phase of 
the wave function whose normalization then explicitly depends (kinematically) on 
$m$.} its dependence is strictly kinematic with gravity only coupling via the 
neutron's de Broglie wavelength, with the COW experiment thus apparently being 
completely compatible with the equivalence principle.  

The author is  extremely indebted to Dr. H. Brown for introducing him to the 
field of gravitationally induced quantum interference, and would like to thank 
him, Dr. W. Moreau, Dr. R. Jones, and Dr. J. Javanainen for many helpful 
discussions. The author would like to thank the University of Canterbury 
in Christchurch, New Zealand for the award of an Erskine fellowship as well as 
for its kind hospitality while much of this work was performed. The author
would also like to thank M. Mannheim for preparing the figures. This work has 
been supported in part by the Department of Energy under grant No. 
DE-FG02-92ER40716.00.

\noindent
{\bf Figure Captions}

\medskip
\noindent
Figure (1). The paths followed by waves in a COW type interferometer.

\medskip
\noindent
Figure (2). The paths followed by waves in a double slit experiment.

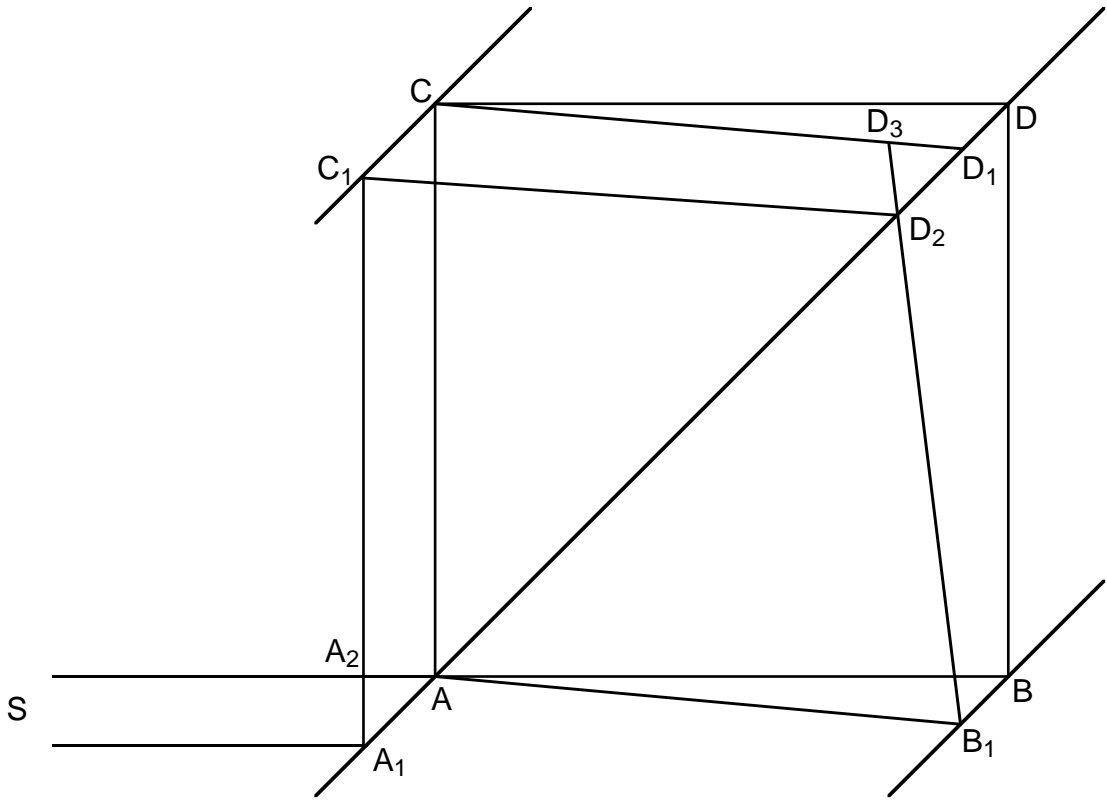
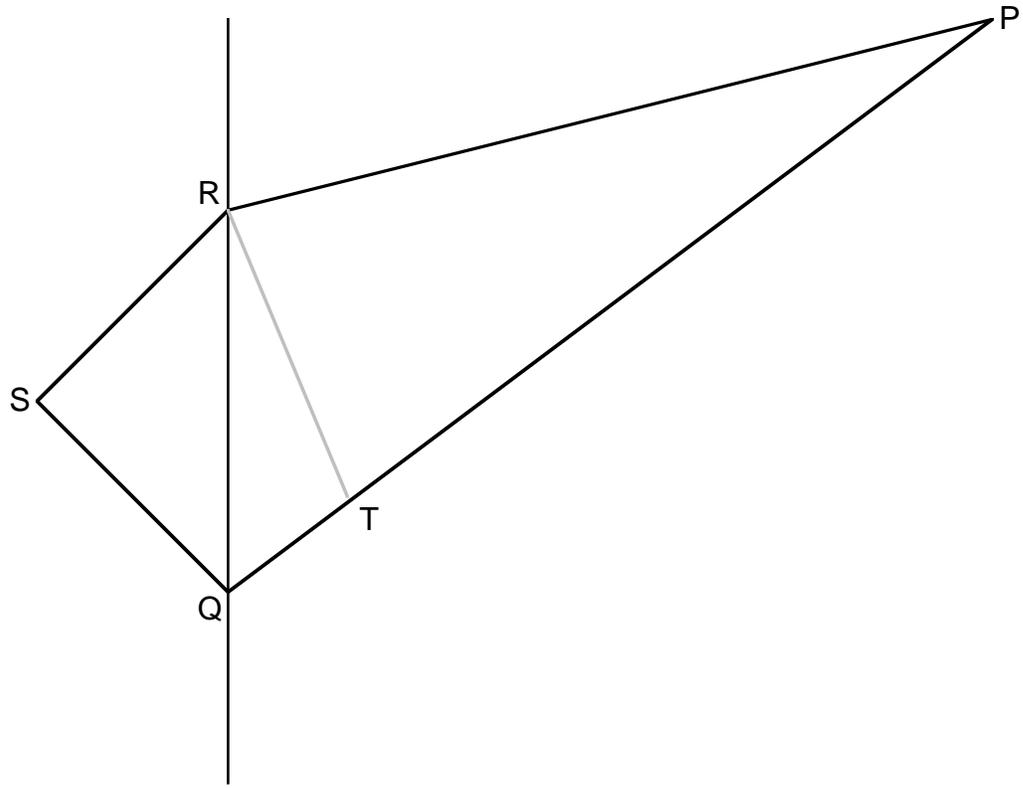
\end{document}